\documentclass[epjST]{svjour}
\usepackage{graphics}

\begin{document}
\title{A multilayer approach for  price dynamics in financial markets}
\author{Alessio Emanuele Biondo\inst{1} \and Alessandro Pluchino\inst{2} \and Andrea Rapisarda\inst{2}\fnmsep\thanks{\email{andrea.rapisarda@ct.infn.it}} }
\institute{ Department of Economics and Business, University of Catania, Italy \and Department of Physics and Astronomy, University of Catania and INFN Sezione  di Catania, Italy.}
\abstract{
   We introduce a new Self-Organized Criticality (SOC) model for simulating price evolution in an artificial financial market, based on a multilayer network of traders. The model also implements, in a quite realistic way with respect to previous studies, the order book dynamics, by considering two assets with variable fundamental prices. Fat tails in the probability distributions of normalized returns are observed, together with other features of real financial markets.  
} 
\maketitle
\section{Introduction}
\label{intro}

The dynamics of  financial markets, with its erratic and irregular behavior at different time scales, 
has stimulated  important theoretical contributions
 by several  physicists and mathematicians 
like Mandelbrot, Stanley, Mantegna, Bouchaud, Farmer, Sornette, Tsallis, \cite{mandelbrot,mantegna-stanley,bouchaud,farmer,sornette,tsallis}, among many others, since long time. 
In particular statistical physics has provided the newborn field of "Econophysics" with new tools and techniques that allow  to model 
and characterize in a quantitative way the apparently unpredictable behavior of price  and trading time dynamics. The recent use of agent-based approaches in financial markets models has also given very useful insights in understanding the often counterintuitive interactions among  heterogeneous agents operating in realistic markets \cite{farmer}.  Recently, herding and imitative behavior among agents has been successfully simulated with Self-Organized-Criticality (SOC) models  and the adoption of random strategies has been shown 
to be an efficient and powerful way to moderate dangerous avalanche effects, diminishing  the occurrence of extreme events \cite{biondo1,biondo2,biondo3,biondo4,biondo5,biondo6}.  
Often these models have adopted  topologies like scale-free and small world networks to describe the social interaction among agents. Such topologies can be further refined for a detailed and realistic description. Very recently multilayer networks have been introduced for a more appropriate framework of several social networks. In this paper we use  a multilayer network, which  as far as we know, has not been used up to now for trading agents, to investigate  price dynamics by means of  an order book based on  two assets.

The paper is organized as follows: in section 2 we  describe the new model; in section 3 simulation results are discussed; in section 4, conclusions are presented.

\section{The Multilayer Network Model}  

\begin{figure}
\centering
\resizebox{0.85\columnwidth}{!}
{  \includegraphics{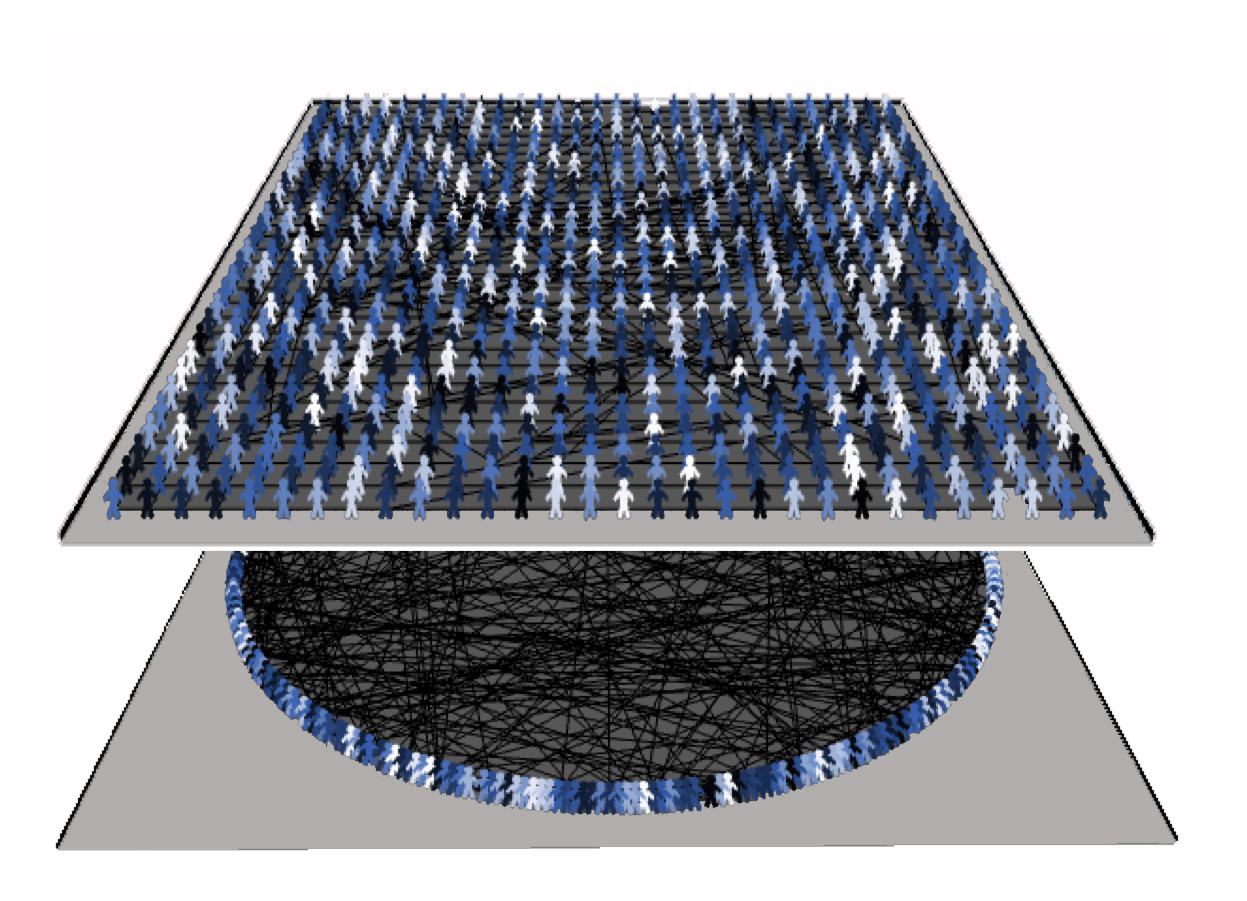} }
\caption{The multilayer network considered in the present model. The top layer is the informative layer, while the bottom one is the trading layer. The nodes in the two layers represent the same traders, but the meaning and the topology of the links are different (see text).}
\label{fig1}      
\end{figure}

In order to simulate the operations of a financial market, the model here presented extends, on one hand, the network framework contained in \cite{biondo5}, where only one (informative) layer was considered, by augmenting it with a second layer, devoted to the order book mechanism. On the other hand, it also extends the single-asset model presented in \cite{biondo6}, by considering a two-assets order book, which makes the trading dynamics more various and interesting. In such a way, we obtain an order-book-driven \textit{Multi-Layer Contagion-Financial-Pricing} model (ML-CFP henceforth), as shown in Fig.\ref{fig1}. Technically, this kind of multilayer network is called {\it multiplex}, since the nodes (traders) are the same in both the layers, changing only the meaning of the edges \cite{boccaletti}.

In brief, the role of the two layers of the ML-CFP model is the following: 

\begin{itemize}
	\item[i)] in the {\bf \em  informative layer}, according to the link configuration given by the network topology, agents collect and share information, therefore deciding their status (bidder, asker or holder) and the (ask or bid) price of their possible orders for the two assets, depending on the global price of the assets at time $t$ and on the herding effect, which induces avalanches of identical investments;
	\item[ii)] in the {\bf \em trading layer}, investors put their orders in the order book, which provides a sort of \textit{compensation room} to execute them, and the next global prices for the two assets emerge from the mutual interaction among all the agents.
\end{itemize}

Let us now explain these features more in detail.

\subsection{The Informative Layer}

The informative layer (the top one in Fig.1) consists of a community of traders $A_i$ (with $i=1,...,N$)  connected among themselves in a two-dimensional square lattice with a \emph{Small World} (SW) topology and with open boundary conditions. The SW network,  introduced in \cite{watts}, is usually adopted  to describe realistic communities in social or economical contexts, thanks to the presence of a small number of long-range links (weak ties) among the regular short-range ones (strong ties). See ref. \cite{biondo2} for more details. 

Each trader in the network is exposed to two flows of information: a {\it global} one and an {\it individual} one \cite{biondo2,biondo3,biondo4,biondo5}:

\begin{itemize}

\item[($a$)] Global flow: this informative pressure reaches all investors uniformly at every time-step, from external sources. Each trader is endowed with a real variable $I_i(t)$ $(i=1,2,...,N)$ that represents her information at time $t$. Initially, at $t=0$, the informative level of each trader is set randomly, in such a way that $I_i(t) \in [0,I_{th}]$, where $I_{th}=1.0$ is a threshold assumed to be the same for all agents. Then, at any time-step $t > 0$, the information accumulated by each trader in her awareness tank is increased by a quantity $\delta I_i$, different for each agent and randomly extracted within the interval $[0,(I_{th}-I_{max}(t))]$. Such an accumulation process may lead a given trader $A_k$, before the other, to exceed her personal threshold value at a given time $t=t_{av}$. In this case, that trader becomes \textit{active} and transmits her opinion, as an informative signal, to her neighbors;

\item[($b$)] Individual flow: it is represented by the opinion spreading among traders, since every one may receive signals from her neighbors, who have possibly passed their threshold. If it happens, it may cause, in turn, that also other agents exceed their thresholds because of this supplementary amount of information, which is additive with regards to the global one ($a$). Such a process explains how the informative cascades may generate herding in the market.
\end{itemize}

The information transfer is realized according to the following simple mechanism, analogous to the energy transmission in earthquake dynamics \cite{caruso1,caruso2}:

\begin{equation}          
I_k > I_{th}  \Rightarrow \left\{ 
	\begin{array}{l}
       I_k \rightarrow 0, \\
       I_{nn} \rightarrow I_{nn} + \frac{\alpha}{N_{nn}} I_k,
       \end{array} 
	\right.
	\label{av_dyn}
\end{equation}

where $nn$ denotes the set of nearest-neighbors of the active agent $A_k$. $N_{nn}$ is the number of direct neighbors, and the parameter $\alpha$ controls the level of dissipation of the information during the dynamics ($\alpha=1$ if there is no dissipation): it is realistic to presume that part of the information content is lost in transmission, therefore in our simulations we always adopted $\alpha<1$.   

\begin{figure}
\centering
\resizebox{0.8\columnwidth}{!}
{  \includegraphics{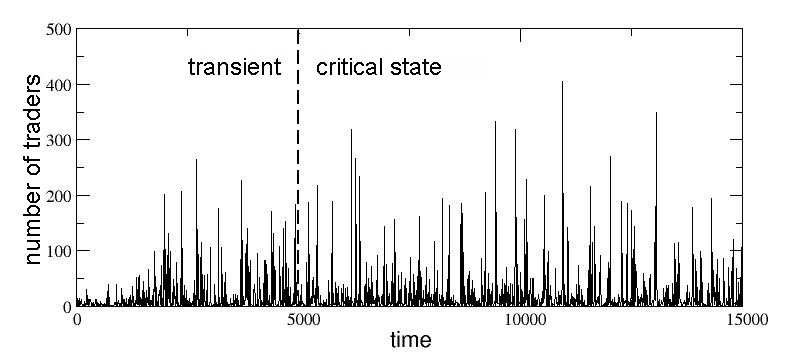} }
\caption{Time series of herding avalanches (see text).}
\label{fig2}      
\end{figure}

When, at a given time $t$, an agent - that we call {\it trigger agent} - overcomes her threshold, i.e. when she reaches a level of knowledge that she considers satisfying, she transmits the (individual) information about her status and, possibly, price for the two assets to her neighbors in the network. 
In turn, the neighbors can overcome their threshold too: in this case they imitate both status and prices of the first agent and transmit the same information to their neighbors following Eq.\ref{av_dyn}, and so on (notice that this is another difference with respect to \cite{biondo6}, where only the status of the first active agent was imitated). In such a way, the herding avalanche can develop. At the end of each avalanche, all the traders involved in the herding process operate in the same way, while the others act independently.   
After a given number of time-steps, such a dynamical rules drive the system into a self-organized critical state, where herding avalanches of every size can happen. In Fig.\ref{fig2} it is shown that a transient of $5000$ time-steps is largely enough for the system to enter into such a critical state.
At this point, the trading layer dynamics come into  action and all the orders are organized in the order book, which operates the matching for the transactions to be actually done. Then, the new price for each of the two assets is determined as explained in the next subsection. 

\subsection{The Trading Layer}

The topology of the trading layer (the bottom one in Fig.1) is that of a fully connected network, since each agent can trade with all the others through the order book dynamics.
In our model we consider an ideal financial market where only two assets exist, each one with its own order book, and where money has an ancillary function, just for transactions regulation. Let us imagine, first, that the trading layer is not influenced by the herding avalanches of the informative layer. In this case we could consider the process of status setting and price formation for an individual agent as independent from the behavior of the other agents.
The traders $A_i$ (with $i=1,...,N$) are endowed, at the beginning of each simulation (i.e. at t=0), with an equally valued portfolio, composed by the same initial quantity of money $M_i(0)=M$ ($\forall i$) and  same initial quantities of the two assets $Q_{1i}(0)=Q_{1}$ and $Q_{2i}(0)=Q_2$ ($\forall i$). 
At each time step, the total wealth of each trader is therefore defined as: $W_i(t)=M_i(t)+Q_{1i}(t)\cdot p_{1}(t)+Q_{2i}(t)\cdot p_{2}(t)$, where $p_{1}(t)$ and $p_{2}(t)$ are the global prices of the two assets at time $t$. At $t=0$, of course, all traders will have the same initial wealth $W_i(0)=M_i(0)+Q_1\cdot p_{1}(0)+Q_2(t)\cdot p_{2}(0)$, being $p_1(0)$ and $p_2(0)$ the initial asset prices.

Two groups of traders do exist in the market: fundamentalists and chartists. At each time-step, traders will behave differently according to their \textit{character}.

{\bf Fundamentalists}: They  presume the existence of a \emph{fundamental value} for each asset, $F_{V_1}$ for asset $1$ and $F_{V_2}$ for asset $2$, and believe that the market prices will always tend to those fundamental values. At variance with \cite{biondo6}, where the fundamental value was fixed at the beginning and did not change in time, here $F_{V_1}$ and $F_{V_2}$ change every $t_f$ time-steps following the rules: 
\begin{equation}
{F_V}_1{(t+t_f)}={{F_V}_1}{(t)}+{D_1}{(t)}
\end{equation}
\begin{equation}
{F_V}_2{(t+t_f)}={{F_V}_2}{(t)} +{D_2}{(t)}
\end{equation}
where ${F_{V_1}}{(0)}=0$ and ${F_{V_2}}{(0)}=0$, while ${D_1}{(t)}$ and ${D_2}{(t)}$ are random variables extracted from two normal distributions with zero mean and standard deviations ${\sigma_{1}}_f$ and ${\sigma_{2}}_f$, respectively. This corresponds to the assumption that, in both cases, dividends follow a random walk. 
The fundamental values are then used by each fundamentalist in order to build a personal opinion about the \textit{correct} prices for the assets, named \textit{fundamental prices}, ${p_{F}}_1(t)$ and ${p_{F}}_2(t)$, being computed as
\begin{equation}
{p_F}_1(t)={p_1}(0) + {F_V}_1(t)+ \Theta
\end{equation}
\begin{equation}
{p_F}_2(t)={p_2}(0) + {F_V}_2(t)+ \Theta
\end{equation}
where $\Theta$ is a parameter randomly chosen in the interval $(-\theta, \theta)$, in order to account for the heterogeneity of investors. Thus, fundamentalists form their expected prices for the two assets according to
\begin{equation}
E[p_1{(t+1)}]=p_1(t)+\phi \cdot [{p_{F}}_1(t) - p_1(t)] + \epsilon
\label{fundamentalists-expectation1}
\end{equation}
\begin{equation}
E[p_2{(t+1)}]=p_2(t)+\phi \cdot [{p_{F}}_2(t) - p_2(t)] + \epsilon
\label{fundamentalists-expectation2}
\end{equation}
where the parameter $\phi$ is a sensitivity parameter that describes the expected speed of convergence to the fundamental prices and $\epsilon$ is a stochastic noise term, randomly chosen in the interval $(-\sigma, \sigma)$. In order to limit the number of parameters, we let the value of $\phi$ be unique and fixed even if, in principle, it could be different for the two assets and for each trader of this group.

{\bf Chartists}: They decide their behavior according to their inspection of past prices. Therefore, before defining their expectation for the future, they will analyze the past dynamics of the two asset price series. In particular, they consider the information coming from such an inspection as two \textit{past reference values} ${P_{RV}}_1(t)$ and ${P_{RV}}_2(t)$, computed at any $t$ by averaging the previous prices over a time window of length $T$, different for each chartist and randomly chosen in the interval $(2,T_{max})$:
\begin{equation}
{P_{RV}}_1(t)=\frac{1}{T} \sum_{j=t-T}^{t} {p_1(j)}
\label{prv1}
\end{equation}
\begin{equation}
{P_{RV}}_2(t)=\frac{1}{T} \sum_{j=t-T}^{t} {p_2(j)}
\label{prv2}
\end{equation}
For sake of simplicity, we consider the same value of $T$ for both the assets. Then, the expected prices for the next time-step are determined by each chartist as
\begin{equation}
E[p_1{(t+1)}]=p_1(t)+\frac{\kappa}{T} \cdot [p_1(t) - {P_{RV}}_1(t) ] + \epsilon
\label{chartists-expectation1}
\end{equation}
\begin{equation}
E[p_2{(t+1)}]=p_2(t)+\frac{\kappa}{T} \cdot [p_2(t) - {P_{RV}}_2(t) ] + \epsilon
\label{chartists-expectation2}
\end{equation}
where $\kappa$ (a constant) is the sensitivity parameter and $\epsilon$ is, again, a stochastic noise term defined as in Eq.\ref{fundamentalists-expectation1} and Eq.\ref{fundamentalists-expectation2}).

In order to choose the status of the traders a sensitivity threshold $\tau$ - common to both the assets - has been introduced in the model, in such a way that, if the expectations are not \textit{sufficiently} strong, i.e. if $p_1(t) - \tau < E[p_1(t+1)] < p_1(t) + \tau$ and $p_2(t) - \tau < E[p_2(t+1)] < p_2(t) + \tau$, the trader will decide to hold on, without setting any order.  On the other hand, if $E[p_1(t+1)] > p_1(t) + \tau$ or $E[p_2(t+1)] > p_2(t) + \tau$ traders will expect a rise in the market price of the corresponding asset and decide to buy, setting their status on {\it bidder}. If, on the contrary, $E[p_1(t+1)] < p_1(t) - \tau$ or $E[p_2(t+1)] < p_2(t) - \tau$, traders will expect a fall in the market price of the corresponding asset and they will decide to sell, setting their status on {\it asker}. Of course, traders who decide to buy must have a positive amount of money ($M_i>0$) and, similarly, those who decide to sell must have a positive amount of the assets (${Q_1}_i>0$ or ${Q_2}_i>0$). 

Once the individual status and about the two assets has been decided, each trader sets her orders in each of the two books by choosing the preferred prices for the transactions. As in \cite{biondo6}, we keep the order mechanism as simple as possible and allow for a maximum of a single order-quantity for each asset. Both in case of sales and purchases, the prices chosen by each trader for the transcription in the order books (personal bid price for bidders and personal ask price for askers) are functions of the expectations that inspired the status of the same trader about the two assets. Since orders are always of quantity $1$, bid (ask) prices are decided by traders by means of simple price setting rules that describe their willingness to pay (to accept) according to their expectations, instead of being defined by means of function optimization procedures. The heterogeneity of traders is embedded in the model by defining feasible intervals from which each investor can extract her bid/ask prices. The rules, that are exactly the same for both the two assets considered independently one from each other, are the following:  

\begin{itemize}
\item[i)] if the status is \textit{bidder}, the chosen bid price will be extracted (with uniform probability) from a range whose minimum and maximum are defined as follows.
    \begin{itemize}
       \item[min:] since it is not convenient for any buyer to set a bid price too low, because no seller would accept to sell, the lower bound for the bid price setting at time $t+1$ is equal to the best ask price (i.e. the lowest one) observed at time $t$;
       \item[max:] since the reason why the investor is bidding is that her expected price is higher than the current one, the upper bound for the bid price setting is exactly that expected price (but, of course, in case the trader has not enough money, the maximum value that she can bid is limited to the owned money);
    \end{itemize} 
\item[ii)] if the status is \textit{asker}, the chosen ask price will be extracted (with uniform probability) from a range, whose minimum and maximum are defined as follows.
    \begin{itemize}
       \item[min:] since the reason why the trader is selling is that her expected price is lower than the current one, the lower bound for the ask price setting is the worst scenario that she infers, i.e. the expected price;
       \item[max:] since it is not convenient for any seller to set an ask price too high, because no buyer would accept to buy, the upper bound for the ask price setting in $t+1$ is the expected price plus $\beta$ times the difference between the current price and the expected one, where $\beta$ is an adjustable parameter chosen in order to balance the number of askers and bidders.
    \end{itemize}
\end{itemize}

After status and price setting activities, orders for a ``$+1$" or ``$-1$" quantity are posted in the two books. 
This is exactly the point where the interaction between the fully connected trading layer and the SW informative one becomes crucial. Actually, as anticipated in the previous subsection, in presence of herding avalanches all the traders involved in the avalanche, regardless of their character (fundamentalist or chartist), will imitate both the status and the price of the agent who started the avalanche itself. And this, of course, strongly influences the order books aspect. 

Once posted all the orders in the two books, considered as independent one from the other, both sides (buy and sell orders) are ranked accordingly with their associated prices. Bid prices are ranked in decreasing order of willingness to pay: in such a way, the trader who has set the highest bid price (namely the \textit{best-bid}) will be the top of the list and will have the priority in transactions. Conversely, ask prices are ranked in increasing order of willingness to accept: the trader with the lowest willingness to accept (who sets the so-called \textit{best-ask}) will be the top of the list and will have the priority in transaction execution. Then, the matching is done by comparing the best ask and the best bid for the two assets. The number of transactions ${N_T}_1$ and ${N_T}_2$that actually does occur for each asset between askers (whose total number is, respectively, ${N_a}_1$ and ${N_a}_2$) and bidders (whose total number is, respectively, ${N_b}_1$ and ${N_b}_2$) strictly depends on such a comparison. Actually, only if best-bid $>$ best-ask we have ${N_T}_1>0$ or ${N_T}_2>0$, i.e. a given number of transactions do occur, depending on the matching among ask and bid prices present in the order book. After the first transaction, occurring among traders who posted their own order at the best price, both from the demand or the supply side, transactions continue following the order in both the books (ascending for the ask list and descending for the bid list) until the bid price is greater than the ask price and all the transactions are regulated at the ask price. Finally, if ${p_L}_1$ and ${p_L}_2$ are, respectively, the ask prices of the last transactions occurred in the two order books, the new global asset prices for the two assets will be determined as

\begin{equation}
{p_1}(t+1) = {p_L}_1 + \delta \cdot \omega_2 
\end{equation}
\begin{equation}
{p_2}(t+1) = {p_L}_2 + \delta \cdot \omega_1 
\end{equation}

\noindent where $\omega_1$ and $\omega_2$ are the market imbalances for the two assets, defined as

\begin{equation}
\left \{
\begin{array}{l}
\omega_1=  {N_b}_1 - {N_T}_1  ~~~~~~~~if~~ ~~ {N_b}_1  \ge {N_a}_1 > 0   \\
\omega_1= -({N_a}_1 - {N_T}_1) ~~~if ~~~~  0 < {N_b}_1 < {N_a}_1 
\end{array} 
\right.
\end{equation}

\begin{equation}
\left\{
\begin{array}{l}
\omega_2=  {N_b}_2 - {N_T}_2  ~~~~~~~~if ~~~~ {N_b}_2  \ge {N_a}_2 > 0   \\
\omega_2= -({N_a}_2 - {N_T}_2) ~~~if ~~~~  0 < {N_b}_2 < {N_a}_2
\end{array} 
\right.
\end{equation}

while $\delta$ is a parameter which quantifies the degree of correlation between the two asset prices.

In such a way, we introduce a feedback mechanism that, according to the unsatisfied side of the market (i.e. either bidders or askers who could not trade for missing counterparts) for a given asset, do influence the price of the other asset, which receives a proportional shift $\delta \cdot \omega$. Thus, for example, in case of an excess of demand for, say, asset $2$ (i.e. bidders are greater in number than askers and therefore some of them cannot trade the asset $2$ at the desired price), the price of asset $1$ will be increased proportionally to the excess itself. Conversely, if askers are greater in number than bidders for asset $2$, the price of asset $1$ is decreased proportionally to the excess of supply. Of course, the same happens by inverting the labels of the two assets.  
  
In the next section we will combine the herding dynamics of the informative layer with the order book mechanism of the trading layer, in order to explore the behavior of the two asset prices through several numerical simulations. 

\begin{figure}
\centering
\resizebox{0.95\columnwidth}{!}
{  \includegraphics{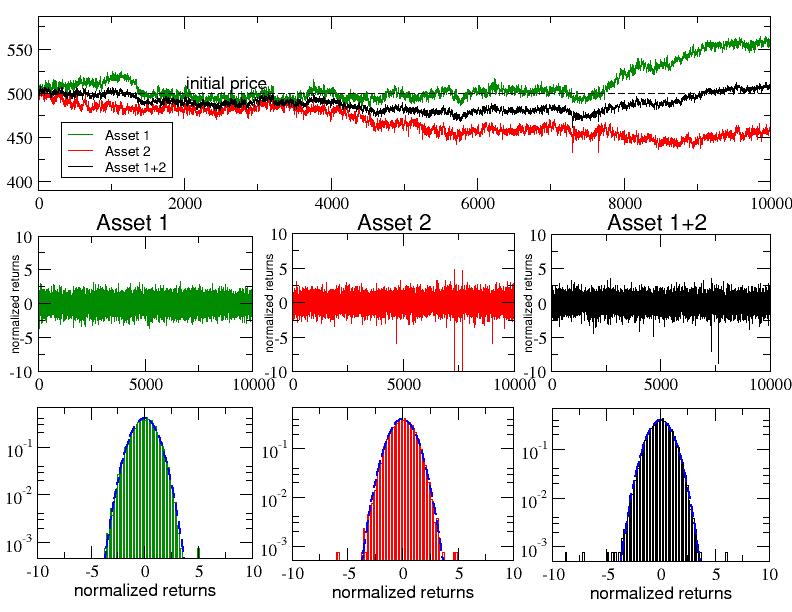} }
\caption{Top panel: typical time series for the global prices of the two assets ($p_1(t)$ and $p_2(t)$) in absence of correlations ($\delta=0.00$) and for the weighted average price $p(t)$ ($1+2$). The initial price is also indicated as an horizontal dashed line. Middle panels: normalized returns of the three price series. Bottom right panel: probability density distributions (PDFs) of the normalized returns compared with Gaussian distributions of unitary variance (dashed lines). }
\label{fig3}      
\end{figure}

\section{Numerical Results }

We present here the numerical results of a typical run of the ML-CFP model, analyzing both its macroscopic and microscopic details, and plotting the final distributions of its main quantities.

We consider a network of $N=900$ traders, with $25\%$ of fundamentalists ($225$ agents) and $75\%$ of chartists ($675$ agents). The (typical) initial setup for the values of the control parameters of the model is the following: $p_1(0)=p_2(0)=500$ (initial asset prices), $\alpha=0.95$ (level of conservation of information), ${\sigma_1}_f={\sigma_2}_f=1$ (standard deviations of the normal distribution for the fundamental values ${F_V}_1(t)$ and ${F_V}_2(t)$), $t_f=10$ (time increment for ${F_V}_1(t)$ and ${F_V}_2(t)$), $\Theta=30$ (range of variation for the fundamentalists' heterogeneity), $\phi=0.5$ (sensitivity parameter for fundamentalists), $T_{max}=100$ (maximum extension of the window for chartists), $\kappa=2.0$ (sensitivity parameter for chartists), $\sigma=30$ (maximum intensity of the stochastic noise for the expectation values), $\tau=15$ (sensitivity threshold for the status setting), $M=40000$ (initial quantity of money) and $Q_1=Q_2=200$ (initial endowment of the two assets).     

\subsection{Uncorrelated Assets}

Let us first take into account the case $\delta=0.00$, i.e. a situation where there is no correlation between the two assets. 

In the top panel of Fig.\ref{fig3} we show a typical time evolution of the two global asset prices $p_1(t)$ and $p_2(t)$, and for the weighted average price $p(t)$ defined as

\begin{equation}
p(t) = p_1(t) \cdot w_1 + p_2(t) \cdot w_2
\end{equation}

where the weights $w_1=Q_1/(Q_1+Q_2)$ and $w_2=Q_2/(Q_1+Q_2)$ are fixed by the initial endowment of the two assets. According to our parameter's choice, $w_1=w_2=0.5$, therefore $p(t)$ is simply the average of $p_1(t)$ and $p_2(t)$. 

In this panel we plot the first $10000$ time-steps after a transient of $5000$ time-steps, starting from the common initial price ($500$), which is highlighted by an horizontal dashed line. Although weakly fluctuating, due to the effect of the herding avalanches, in absence of mutual correlations the two asset prices follow distinct time evolutions. In this case, the volatility of both the prices, as well as that one of the averaged price,  remain almost normal. This can be shown by plotting the normalized returns of the prices, generally defined as \mbox{$r^{norm}_{t}=(r_{t} - r_{av})/r_{stdev}$}, where $r_{t}=log(p_{t+1})-log(p_{t})$ are the logaritmic returns while $ r_{av}$ and $ r_{stdev}$ are, respectively, their mean and standard deviation calculated over the whole time series. In the middle panels we report the three time-series while in the bottom panels their corresponding PDFs. The Gaussian shape of the three curves is evident if compared with normal distributions with unitary variance, also reported as dashed lines. This means that, if the asset prices are not correlated, one does not observe extreme events in the market fluctuations, since the system evidently self-organizes, maintaining a dynamical balance between purchase orders and sales.   

\begin{figure}
\centering
\resizebox{0.95\columnwidth}{!}
{  \includegraphics{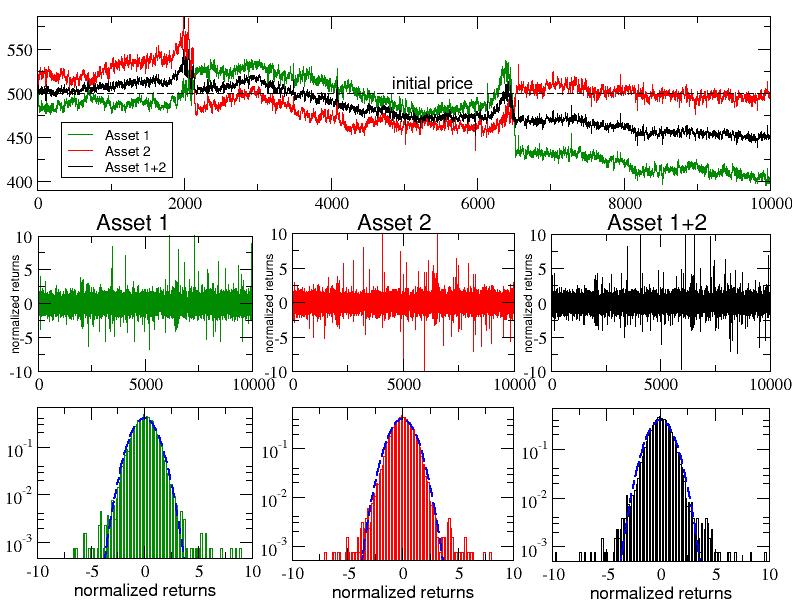} }
\caption{Top panel: typical time series for the global prices of the two assets ($p_1(t)$ and $p_2(t)$) and for their average $p(t)$ in presence of correlations ($\delta=0.03$). The initial price is, again, indicated as an horizontal dashed line. Middle panels: normalized returns of the three price series. Bottom right panel: probability density distributions (PDFs) of the normalized returns compared with Gaussian distributions of unitary variance (dashed lines). }
\label{fig4}      
\end{figure}

\begin{figure}
\centering
\resizebox{0.6\columnwidth}{!}
{  \includegraphics{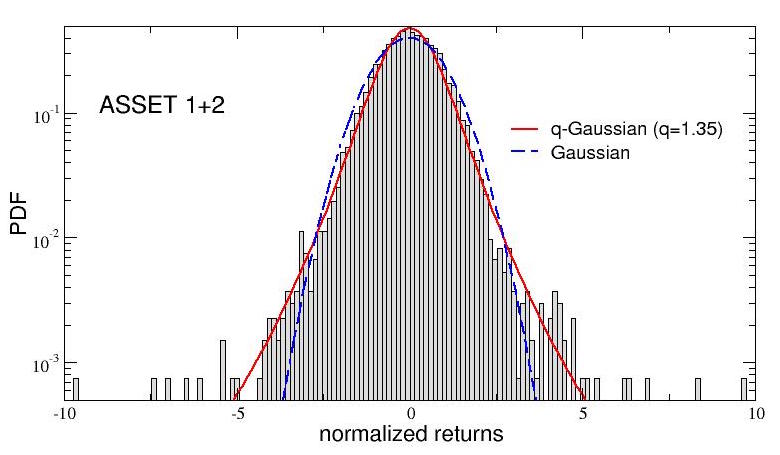} }
\caption{The probability distribution of the normalized returns for the average price $p(t)$ (asset $1+2$) can be well fitted by a fat tailed q-Gaussian curve with $q=1.35$.  }
\label{fig5}      
\end{figure}

\subsection{Correlated Assets}

Quite different is the situation if one consider an even though weak correlation between the two assets, by setting $\delta=0.03$ in Eqs.12 and 13.  
        
In Fig.\ref{fig4} we show the same plots than in Fig.3, but now the fluctuations of the two prices $p_1(t)$ and $p_2(t)$ are visibly much stronger than before. Furthermore, their time evolutions (top panel) appear to be strongly coupled: reversals in the price values can be observed at $2000$ and $6500$ time-steps, where sudden price falls of, respectively, $p_2$ and $p_1$, take place. Such a dynamics, due again to the presence of herding avalanches but also to the newly introduced prices correlation, induces in turn a higher volatility, as shown in the time series of the middle panels. As a consequence, fat tails start to appear in the corresponding PDFs of the bottom panels, thus confirming the presence of extreme financial events. 

In Fig.\ref{fig5}, an enlargement of the PDF for the normalized returns of the average price $p_t$ (asset $1+2$) is reported: deviations from the normal behavior (dashed curve) are clearly visible and the fat tailed shape can be quite well fitted by a q-Gaussian curve (full line) with $q=1.35$. The latter is a fat tailed distribution, typical of non-extensive statistical mechanics, defined as $y=A(1-(1-q) \beta x^2)^{1/(1-q)}$, where the entropic index $q$ measures the deviation from gaussian behavior (for $q=1$ the Gaussian shape is recovered). 

\begin{figure}
\centering
\resizebox{0.95\columnwidth}{!}
{  \includegraphics{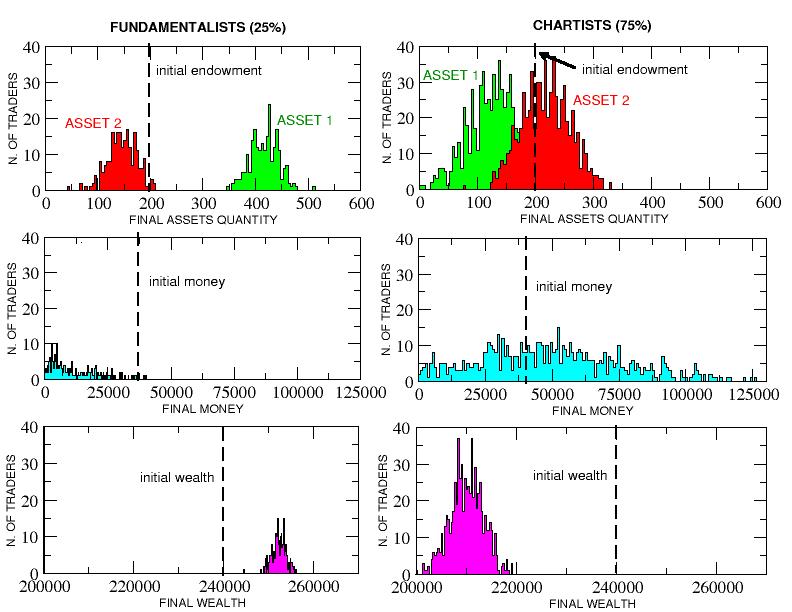} }
\caption{fFinal distributions of asset quantity, money and wealth for, respectively, fundamentalists (left column) and chartists (right column). The initial values of the three quantities, equal for all the traders, are also reported as dashed vertical lines. }
\label{fig6}      
\end{figure}

Let us now look to the microscopic details of some interesting quantities as they appear at the end of the simulation. In Fig.\ref{fig6}, the final distributions of asset quantity, money and wealth for, respectively, fundamentalists (left column) and chartists (right column), are plotted. The initial values of the three quantities, equal for all the traders, are also reported as dashed vertical lines. As one could see, fundamentalists accumulate a great quantity of asset $1$, while mainly tend to sell asset $2$: as a consequence, at the end of the simulation all of them have less money with respect to the beginning, but their total wealth stay always above the initial value. On the other hand, chartists mainly tend to sell asset $1$, while have a quite neutral behavior with respect to asset $2$: in such a way, many of them increase their initial capital in terms of money, even if their total wealth always remain well below the initial value. Such a scenario is consistent with the details about the average percentage of fundamentalists and chartists who buy or sell, calculated over the whole simulation. In this respect, at a first sight, the situation does appear quite equilibrated: actually, for fundamentalists, we have $26\%$  of buyers and $24\%$ of sellers, while, for chartists, $25\%$ of buyers and $26\%$ of sellers. However, although very small, in the long term these slight discrepancies account for the different attitude of the two kind of traders and, in turn, for their different wealth and portfolio.               

The features observed for this typical run are quite robust and remain substantially unchanged if one varies not only the relative proportion of fundamentalists and chartists, but also the initial value of the asset price, the initial asset endowment or the initial money quantity. On the other hand, they are quite sensitive with respect to variations in some control parameters, like the sensitivity for expectation prices or the sensitivity threshold for the status setting: by changing these parameters, the previously observed quite good equilibrium between bidders and askers becomes much more unstable and, typically, one of the two trading groups, fundamentalists or chartists, start to buy the asset much more than the other one, thus generating a spiral effect that leads fundamentalists or chartists to spend all its money thus taking, in fact, out of the market.

\section{Conclusions}
\label{sec:3}

We have presented a new model of order book, the ML-CFP model, able to describe price dynamics in a financial market with two assets, realized through a multilayer network of heterogeneous agents. This realistic framework produce  interesting numerical results, which - despite the simplifying assumptions about assets and orders - adhere to some typical features of real financial markets. 
Further numerical explorations of this model, as for example a more detailed parametric analysis and the influence of random traders  are in progress, but  will be reported elsewhere. 

\section{Acknowledgements}
\label{sec:4}

We would like to dedicate this paper to Alberto Robledo for his 70th birthday. One of us (A.R.) would like to thank the organizers of the conference in honor of Alberto Robledo for the warm hospitality and the financial support.
This study was partially supported by the FIR Research Project 2014 N.ABDD94 of the University of Catania.

\end{document}